# Magnetic order dependent photoluminescence from high energy excitons in hBN protected few-layer CrSBr

Xiaohua Wu [1,‡], Junyang Chen [1,‡], Mingqiang Gu [1,‡], Yujun Zhang [3], Shanmin Wang [1,2], Yanan Dai [1,2,4], Qihang Liu [1,2,4], Yue Zhao [1,2], and Mingyuan Huang [1,2, *]

[1] Department of Physics, State Key Laboratory of Quantum Functional Materials, and Guangdong Basic Research Center of Excellence for Quantum Science, Southern University of Science and Technology, Shenzhen 518055, China

[2] Guangdong Provincial Key Laboratory of Advanced Thermoelectric Materials and Device Physics, Southern University of Science and Technology, Shenzhen 518055, China

[3] School of Physics and Astronomy and Key Lab of Quantum Information of Yunnan Province, Yunnan University, Kunming 650091, China

[4] Quantum Science Center of Guangdong–Hong Kong–Macao Greater Bay Area (Guangdong), Shenzhen 518045, China

‡These authors contributed equally to this work

Contact author: huangmy@sustech.edu.cn

**Abstract**

The detection and manipulation of the spin configurations in layered magnetic semiconductors hold significant interest for developing spintronic devices in two-dimensional limit. In this letter, we report a systematical study on the photoluminescence (PL) from the high energy excitons in few-layer CrSBr and its application on detecting the spin configurations. Besides the broad excitonic emission peak ($X_l$) at around 1.34 eV, we also observed another strong excitonic emission peak ($X_h$) at around 1.37 eV in hBN encapsulated 2L sample, which splits into two peaks in 3L and 4L samples. With help of the first principles calculations, we conclude that the $X_h$ peak is associated with the transition between the top valence band and the second lowest conduction band, which is forbidden by the inversion symmetry in 1L CrSBr. Furthermore, the position and intensity of the $X_h$ peak are strongly dependent on the interlayer magnetic order of the CrSBr samples, which provides an efficient way to probe their spin configurations. In addition, when the magnetic field is applied at the easy axis direction, we resolve an intermediate magnetic state besides the antiferromagnetic and ferromagnetic states in 3L and 4L samples. Our results reveal few-layer CrSBr as an ideal platform to study the interaction between the excitons and magnetism.

KEYWORDS: 2D magnets, symmetry breaking, exciton emission, CrSBr, magneto-optical effect

The discovery of intrinsic ferromagnetic (FM) order down to monolayer has stimulated intensive research interests in both fundamental physics and potential spintronics applications in the 2D limit[1–9]. The magneto-optical properties, magnetic excitations and manipulations of magnetic order by pressure and electrostatic doping in van der Waals magnetic materials have been extensively explored[10–14]. Recently, antiferromagnetic (AFM) semiconductor CrSBr with intralayer FM order along the *b*-axis and AFM interlayer coupling, has attracted a lot of attention due to its magnetic properties and relatively high Néel temperature[15–31]. More interestingly, CrSBr demonstrates plenty of novel magneto-optical phenomena, including anisotropic exciton emission coupled with magnetic order[18,32], magnon-exciton coupling[33–35], exciton-polaritons[36,37]. However, most of the experiments are done in CrSBr samples without protection, and the optical properties of hBN protected CrSBr are still elusive.

In this letter, we present a systematical study on the optical properties of hBN protected few-layer CrSBr to investigate the correlation between the exciton emission and spin configuration. By employing cryogenic photoluminescence (PL) spectroscopy, we observe a bright high energy exciton peak ($X_h$) at around 1.37 eV in hBN encapsulated 2L CrSBr sample, besides a broad exciton peak ($X_l$) at around 1.34 eV which has been reported in the previous literatures[18,30]. This peak splits into two peaks in 3L and 4L samples, which are associated with excitons from surface and inside layers. According to the first principles calculations, the $X_h$ exciton is from the transition between the second lowest conduction band and the top valence band, which is forbidden by the inversion symmetry in 1L CrSBr. By applying inhomogeneous strain in 1L sample, the spatial inversion symmetry is broken and the $X_h$ exciton turns bright. In addition, through the magnetic field-dependent experiments, strong magneto-optical coupling has been observed for the $X_h$ exciton, which enable us to probe the spin configuration easily by PL or differential reflectance spectrum. Furthermore, we also discover intermediate magnetic states beyond the normal AFM or FM states in 3L and 4L samples, when the magnetic field is applied along the easy axis.

Fig. 1a displays the photoluminescence (PL) spectra of 1-4L CrSBr samples which are protected by hBN encapsulation at 1.7 K. The PL spectrum of 1L sample only shows one asymmetrical and broad peak at around 1.34 eV labeled as $X_l$ which is originated from the lowest energy excitonic transition and evolved into a more complex feature in 2-4L samples, and these results are consistent with previous reports[27,30,31,38]. In a sharp contrast, we observe a new narrow and strong excitonic emission peak with width of ~3 meV at about 1.37 eV labeled as $X_h$ in 2L sample, which splits into two peaks with a separation of ~ 5 meV labeled as $X_h'$ and $X_h$ in 3L and 4L samples. To confirm our observations, the reflectance measurements are performed and the results are presented in Fig. 1b. The high energy excitonic transitions can be well resolved in the differential reflectance spectra, which shows an excellent agreement with the PL spectra. In addition, the new observed excitonic transitions become the dominate features in the reflectance spectra for 3 and 4L samples. Furthermore, the polarization-dependent PL measurements confirm that the high energy excitonic emissions display the same polarization with the one at 1.34 eV (see supplementary Fig. S2 ), which is consistent with the optical anisotropy of CrSBr reported before[17,18,21,39].

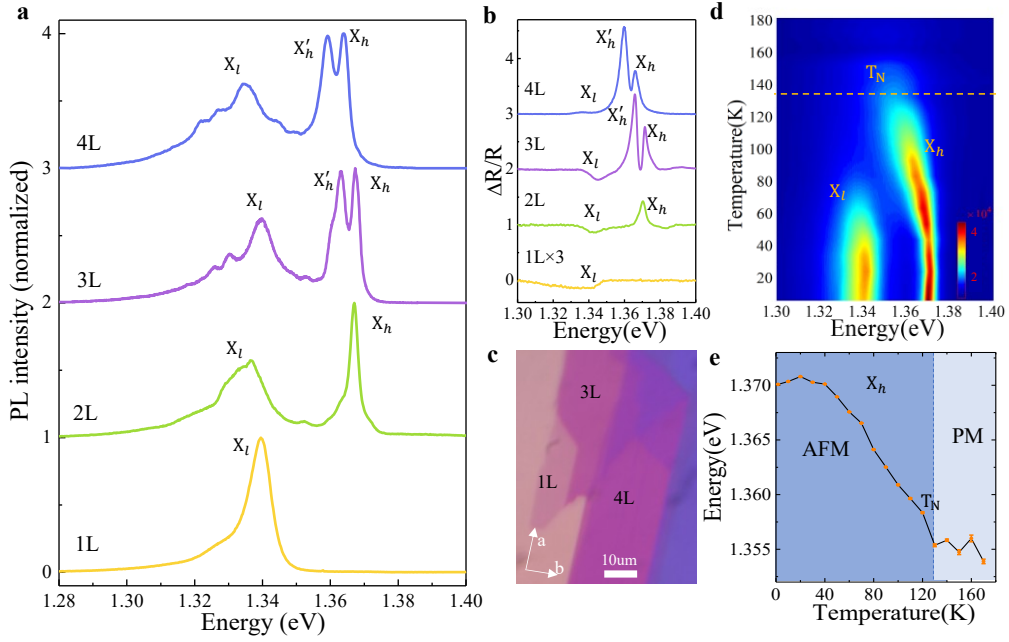

**Fig. 1| Optical properties of hBN protected few-layer CrSBr.** (a) Photoluminescence (PL) spectra of 1-4L CrSBr. The spectra are normalized with maximum intensity. (b) Differential reflectance spectra of 1-4L CrSBr. 1L data are magnified threefold for clarity. (c) Optical image of exfoliated few-layer CrSBr. (d) PL spectra of 2L CrSBr as a function of temperature. The golden dashed line represents $T_N$ (Néel temperature) for 2L CrSBr. (e) Temperature dependence of the $X_h$ peak position. Dark blue highlights antiferromagnetic regions ($T < T_N$), while light blue indicates paramagnetic regions.

We next investigated the temperature dependence of the PL spectra of few-layer CrSBr samples. A representative result of 2L CrSBr sample is shown in Fig. 1d (the rest results are provided in the Supplementary Fig. S3). As the temperature increasing, the $X_l$ peak shows slight blueshift and loses its intensity at around 80 K; the $X_h$ peak displays a minor blueshift below 40 K and then a clear redshift until the temperature approaching the Néel temperature ($T_N$) of 140 K. The peak positions of $X_h$ are extracted by Lorentzian fitting and plotted in Figure 1e. In addition, the $X_h$ peak shows clear broadening and intensity dropping above 40 K, and becomes even more broad and weak above $T_N$.

Previous studies have demonstrated that the low energy exciton $X_l$ in few-layer CrSBr is sensitive to the magnetic order[18,38]. To investigate the properties of the high energy excitons ($X'_h$, $X_h$) discovered in hBN protected CrSBr samples, we performed magneto-optical PL studies. Figs. 2a, b and c show the evolution of the PL spectra for 2L CrSBr sample with the magnetic field along the *a*-, *b*- and *c*-axes respectively. From the Fig. 2a, the high energy exciton $X_h$ displays symmetric redshifts from 0 T toward both increasing and decreasing magnetic field, and saturates above 0.5 T and below -0.5 T, which is consistent with the previous differential reflectance measurements[31]. This result can be understood by considering the magnetic order from AFM state at 0 T, to spin canting, and then FM state at high magnetic fields, as indicated by the arrows in the left of Fig. 2a. Besides the red-shifts of the peak positions, we also observe that the $X_h$ exciton undergoes a significant intensity dropping without obvious change of the full width at half maximum (FWHM) from the AFM toward FM state in a sharp contrast to the $X_l$ exciton. A similar dispersive behavior can be observed for the magnetic field along the *c*-axis with a higher saturation field of 1.7 T, as shown Fig. 2c. However, when the magnetic field is applied along the easy axis (*b*-axis), sudden changes can be resolved in the PL spectra at +/-0.2 T as shown in Fig. 2b, which is consistent with the spin-flip transition from the AFM to FM state. Finally, two representative PL spectra from both AFM and FM states are selected and plotted in Fig. 2d, and the peak position redshift and intensity dropping can be clearly resolved.

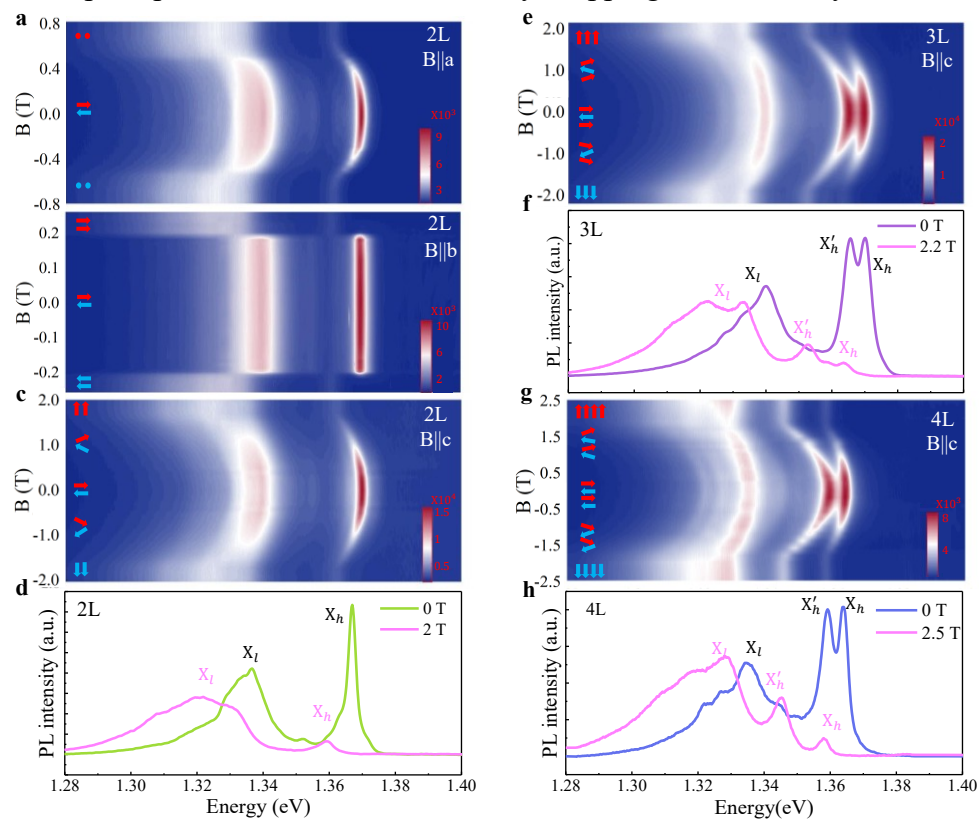

**Fig. 2| PL spectra of few-layer CrSBr under magnetic fields.** (a-c) PL spectra of 2L CrSBr under magnetic fields along the *a*-axis (a), *b*-axis (b), and *c*-axis (c), respectively. Spin orientation is illustrated on the left side. (d) PL spectra of 2L CrSBr at the AFM and FM states. (e, g) PL spectra of 3L(e) and 4L(g) CrSBr under magnetic fields along the *c*-axis. (f, h) PL spectra of 3L (f) and 4L (h) CrSBr at AFM and FM states.

To further investigate the magneto-dispersion in 3L and 4L CrSBr samples, the magneto-optical PL experiments are performed with the magnetic field along the *c*-axis, and the results are plotted in Figs 2e, g with the representative PL spectra from both AFM and FM states in Figs. 2f, h, respectively (the results for the magnetic field along the *a*-axis are provided in the Supplementary Fig. S4). The $X_h$ excitons in 3L and 4L samples display similar energy dispersion from the AFM to FM state with 2L sample, while the curvatures of dispersion for the $X'_h$ excitons are bigger which leads a wider separation between $X'_h$ and $X_h$ excitons at the saturated FM state, as shown in Fig. 2g, h. In addition, both $X'_h$ and $X_h$ excitons undergo similar intensity decreasing with the $X_h$ exciton in 2L sample.

To reveal the origin of the high energy excitons in few-layer CrSBr samples, we performed the first principles GW calculations (details in Methods) to obtain the electronic band structure of 2L CrSBr in the AFM state, as shown in Fig. 3a. Due to the product symmetry of the time-reversal and spatial inversion, the band structure is degenerate in spins which is almost the same with that of 1L CrSBr except the spin degeneracy. According to our calculations, the $X_l$ exciton is assigned as the transition between the top valence band (v) and the lowest conduction band (c1) at around the Γ point, which is dipole-allowed transition; the $X_h$ exciton is assigned as the transition between the top valence band and the second lowest conduction band (c2). From our analysis, the transition v → $c_2$ is dipole-forbidden in 1L CrSBr due to the spatial inversion symmetry, while this transition becomes dipole-allowed in 2L CrSBr because the AFM order breaks the inversion symmetry, as shown in Fig. 3c. In addition, the $X_h'$ excitons in 3L and 4L samples should come from the splitting of the $c_2$ band.

Based on our analysis, the $X_h$ should also exist in 1L CrSBr as dark excitons due to the inversion symmetry. To confirm our theory, we fabricated a 1L CrSBr sample on $SiO_2$/Si substrate with a $SiO_2$ nanoparticle (diameter is about 1um) underneath the sample, as shown in the inset of Fig. 3d. The adhesion between the sample and substrate will produce inhomogeneous strain in CrSBr around the nanoparticle, which breaks its spatial inversion symmetry. The PL spectra of the 1L CrSBr sample are measured near and away from the nanoparticle and plotted in Fig. 3d. Obviously, the $X_h$ becomes a bright exciton after the inversion symmetry breaking. This result clarified the conflictive symmetry assignments of the $c_1$ and $c_2$ bands[18,31], and demonstrated that the $X_h$ exciton emission of 1L CrSBr can be modulated by controlling its symmetry. For example, by fabricating hBN protected CrSBr/$CrCl_3$ heterojunction, the $X_h$ exciton emission of 1L CrSBr can also be observed due to inversion symmetry breaking by the different interface interactions of CrSBr/$CrCl_3$ and hBN/CrSBr (see Supplementary Fig. S5). In addition, for the unprotected few-layer CrSBr samples, the high energy exciton emission can only be observed in 3L and more layered samples, and the high energy exciton was assigned as a bulk exciton in some literatures[40]. However, our results demonstrate that the high exciton exists down to 1L CrSBr.

For few-layer CrSBr, the intensity of the $X_h$ exciton in the PL spectrum is similar with that of the $X_l$ exciton, which is unusual because the $c_1$ band is much more populated than the $c_2$ band at low temperatures after excitation. To understand this phenomenon, the details of the $c_1$ and $c_2$ bands near the Γ point are magnified and plotted in Fig. 3b, which shows that the minima of the $c_2$ band is right at the Γ point and a direct optical transition can happen between the $c_2$ band and the v band, while the minima of the $c_1$ band is a little away from the Γ point and the transition between the $c_1$ band and the v band is mostly indirect. Given the broadness of the $X_l$ peak in the PL spectrum, it might be partially from indirect transition and partially from direct transition. In addition, the strong PL intensity of the $X_h$ exciton in our protected CrSBr samples also requires a relative slow relaxation rate from the $c_2$ band to the $c_1$ band, while the $X_h$ exciton is undetectable in the unprotected 2L CrSBr sample, which might be resulted from the increased relaxation rate after exposed in the air (see Supplementary Fig. S1).

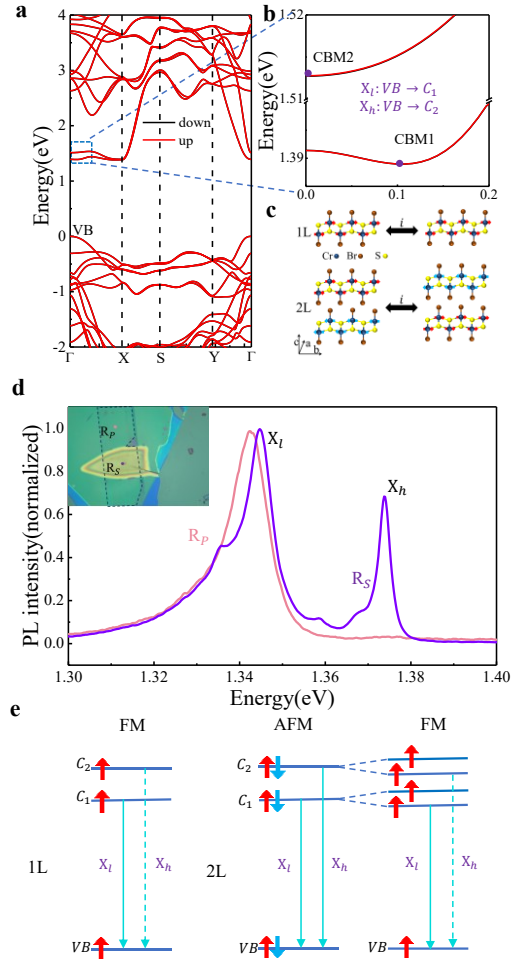

**Fig. 3|** (a) The electronic band structure of 2L CrSBr calculated by using the GW method. (b) Enlarged band structure of two conduction bands near the Γ point, highlighting the conduction band minima of the first (CBM1) and second (CBM2) lowest conduction bands. (c) The spatial inversion symmetry in 1L and 2L CrSBr. (d) The PL spectra of the intrinsic and imhomogeously strained 1L CrSBr. The inset displays the optical image of the strained 1L CrSBr sample with marks of measurement positions. (e) Optical transitions of 1L and 2L CrSBr in AFM and FM states. The solid lines with arrows denote dipole-allowed transitions, while the dashed lines with arrows indicate the dipole-forbidden transitions.

Now we turn to understand the PL spectrum in few-layer CrSBr under magnetic field. The relevant energy bands and corresponding transitions for 1L and 2L CrSBr are simplified and summarized in Fig. 3e. For 2L CrSBr, the $c_1$ and $c_2$ bands split into 2 sub-bands after changing from the AFM to FM state by the applied magnetic field, and this band splitting reduces the bandgaps for v → $c_1$ and v → $c_2$. Consequently, both $X_l$ and $X_h$ excitons display redshifts during the transition. Moreover, the FM order doesn't break its inversion symmetry and the transition v → $c_2$ becomes dipole-forbidden in the FM state, which is consistent with the significant intensity decreasing of the $X_h$ exciton as we observed. Furthermore, the band splitting will decrease the separation between the $c_2$ and $c_1$ sub-bands, and hence increase the relaxation rate of $c_2$ → $c_1$, which might be another important reason for the intensity decreasing of $X'_h$ and $X_h$. In addition, the $X_h$ exciton is still observable in the FM state as shown in Fig. 2d, which means that the inversion symmetry breaking might be not fully originated from the AFM magnetic order in 2L CrSBr, and the interlayer mechanical coupling could be another possible reason. By using the same simple layered model, the inversion symmetry should be preserved for both AFM and FM states in 3L CrSBr, while the split high excitons display significant intensity in the AFM state and similar intensity decreasing in the FM state as shown in Fig. 2f. To fully understand this result might need more sophisticated model by considering the magneto-elastic and interlayer mechanical couplings.

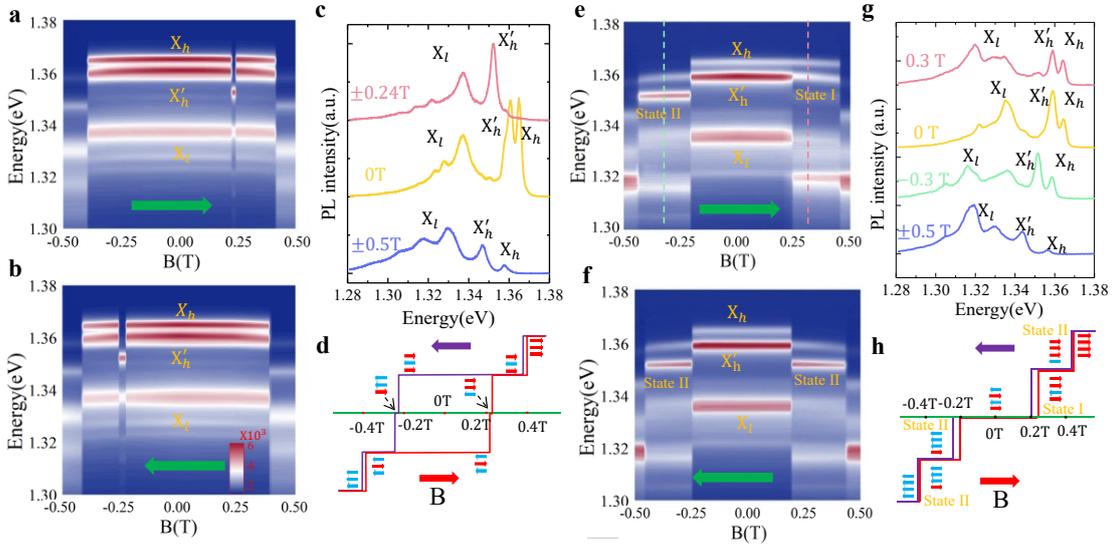

**Fig. 4| The intermediate magnetic states in 3L and 4L CrSBr.** (a-b) PL of 3L CrSBr under the increasing (a) and decreasing (b) magnetic field along the *b*-axis. The green arrows denote the direction of the magnetic field sweeping. (c) PL spectra of 3L CrSBr at 0, 0.24 and 0.5 T (increasing magnetic field). (d) Magnetic order of 3L CrSBr during the magnetic field sweeping cycle along the *b*-axis. (e-f) PL of 4L CrSBr under the increasing (e) and decreasing (f) magnetic field along the *b*-axis. (g) PL spectra of 4L CrSBr at -0.3, 0, 0.3 and 0.5 T (increasing magnetic field). (h) Magnetic order of 4L CrSBr during the magnetic field sweeping cycle along the *b*-axis.

For 3L CrSBr, similar abrupt AFM-FM transitions with larger critical field of ~0.4 T are observed as 2L CrSBr when sweeping the magnetic field along the easy axis (*b*-axis) upward and downward as shown in Figs. 4a and 4b, and a large hysteresis can be found for this spin-flip transition which is different from the situations of magnetic field along the *a*- and *c*-axis. Surprisingly, we also observe an additional transition into an intermediate state at around 0.2 T (upward) and -0.2 T (downward), which is consistent with the previous differential reflectance measurement.[31] The typical PL spectra for the AFM, intermediate and FM states are plotted in Fig. 4c. For the intermediate state, both

$X_h'$ and $X_h$ excitons switch to lower energies, and the $X_h$ losses most of its intensity, while the intensity of $X_h'$ remains strong, which is different from that of the FM state. To understand the spin-flip transitions as a function of magnetic field, we take the upward sweeping direction as an example. Initially, 3L CrSBr is polarized to FM state with spins of all 3 layers toward left below -0.45 T. With increasing the magnetic field, the spins of the middle layer flip to right and forming the AFM order, which is persistent beyond 0 T. Subsequently, the spins of the top (or bottom) layer flip to the right and the sample enters the metastable intermediate state which is energetically unfavorable comparing to the AFM state with the spins of the middle layer toward left and others toward right. Therefore, the sample soon switches back to the AFM state with the spins of the other 2 layers flipping, and finally is polarized to FM state with all spins toward right above 0.45 T. In addition, those spin-flip transitions are summarized in Fig. 4d.

To further explore the metastable intermediate state in few-layer CrSBr, we performed the same PL measurements in 4L samples and a representative result was displayed in Figs. 4e, f.[30,41] For increasing magnetic field in Fig. 4e, the AFM state from -0.2 T to 0.25 T and FM state below -0.44 T and above 0.45 T can be easily identified, and transitional states can be found between the FM and AFM states (-0.44 ~ -0.2 T, 0.25 ~ 0.45 T), which is similar with 4L CrI$_3$. However, two transitional states display different features in their PL spectra: both $X_h'$ and $X_h$ excitons jump to lower energies with minor intensity decreasing at the negative field, while only the $X_h'$ shows intensity decreasing without energy shift for both excitons at the positive field, as shown in Fig. 4g. For the transitional states, one spin (representing one layer) points one direction and the others point the opposite direction, and there are two different choices for the first spin: the inside layer (labeled as state I) and the surface layer (state II). Here, the transitional state at the negative field is assigned as state II by considering that its PL spectrum is similar with that from the intermediate state in 3L sample, and the one at the positive field as state I. For decreasing magnetic field in Fig. 4f, both transitional states are state II, and the spin-flip transitions are summarized in Fig. 4h. By considering the interlayer AFM coupling, state I is energetically more favorable than state II, and direct transition from state II to state I can be observed in some samples, as shown in Supplementary Fig. S7.

In conclusion, we have systematically studied the high energy exciton $X_h$ emission originated from the transition between the second lowest conduction band and the top valence band in hBN protected few-layer CrSBr samples. In the PL spectrum, the $X_h$ exciton is featured as a sharp peak with significant intensity in 2L CrSBr and split into two peaks in 3L and more layered samples. This $X_h$ exciton emission is forbidden by the inversion symmetry in 1L CrSBr and becomes observable in unevenly strained sample. By tuning the AFM order to FM order in CrSBr samples, the $X_h$ exciton displays significant redshift and intensity decreasing, which makes PL spectroscopy as an efficient tool for probing the spin configurations. Furthermore, intermediate magnetic states are resolved besides the normal AFM and FM states in 3L and 4L samples when the magnetic field is applied along the easy axis, and these metastable magnetic states might be used in novel spintronic devices in the future.

## METHODS

### Synthesis of CrSBr crystals and preparation of CrSBr flakes

Single crystals of CrSBr were synthesized by the chemical vapor transport method using the recipe described in reference[42]. Atomically thin CrSBr flakes were mechanically exfoliated onto 285 nm $SiO_2$/Si substrates in an argon-filled glove box, and the sample thickness was first estimated by optical contrast and further confirmed by PL spectroscopy at 1.7 K. The few-layer CrSBr were encapsulated by hBN flakes using the dry transfer technique in the glove box to prevent oxygen and moisture. In strained 1L CrSBr experiments, the silica nanoparticles (NPs) with a particle size of ~1 μm were dispersed in pure ethanol at concentrations of 0.1 g/L and sonicated for 15 min. Next, we use a glass capillary tube to transfer the NPs diluent onto the silicon wafer. Afterward, we used all-dry transfer method to transfer the single-layer hBN encapsulated CrSBr flakes onto NPs.

### Cryogenic PL and reflectance spectroscopy

Both PL and differential reflectance measurements were performed in a closed-cycle cryostat (attoDRY 2100) with 9/3T superconducting vector magnets with a base temperature of 1.7 K. For magneto-measurements along the *a* and *b* axes, the sample holder was mounted on a rotator. For PL measurements, a 532 nm laser with a power of 100 μW was focused onto the samples by using a 40X objective. The backscattering light was collected by the same objective and captured by a spectrometer with a 600-groove/mm grating and a nitrogen-cooled Si CCD. For differential reflection spectroscopy, a tungsten-halogen lamp replaces the 532 nm laser.

### First principles calculations

Density functional theory (DFT) simulation was performed using the Vienna Ab-initio Simulation Package (VASP) code[43] to understand the electronic band structure of CrSBr with different layers. The projector augmented wave (PAW) method[44] was used to treat the core and valence electrons using the following electronic configuration: $3d^54s^1$ for Cr, $3s^23p^4$ for S and $4s^24p^5$ for Br. The Perdew–Burke–Ernzerhof (PBE)-type exchange-correlation functionals (PBEsol) were used in our calculation[45]. The DFT-D3 method of Grimme with zero-damping function was used to include the vdW correction[46]. A Γ-centered fine k-point grid with the resolution of 0.02 (in unit of 2π/Å) is used. For the structure optimization, the atomic positions were relaxed until the forces on each atom were less than 1 meV/Å. The partially self-consistent calculations[47] $GW_0$ was used to consider the band gap correction due to quasiparticle contribution.

## ASSOCIATED CONTENT

Supporting Information
The Supporting Information is available free of charge at

## Notes

The authors declare no competing financial interest.

**Author contributions**
The manuscript was written through contributions of all authors. All authors have given approval to the final version of the manuscript. X.W, J.C and M.G contributed equally. M.H. conceived and designed the research project. S.W. provided hBN crystals. X.W. and J.C. fabricated the sample devices and performed the experiments. M.G. and Q.L. performed the theoretical analysis and DFT calculations. The manuscript was prepared by X.W., J.C., M.G., Y.D. and M.H. in consultation with all other authors. All authors read and commented on the manuscript.

**Acknowledgements**
This Work was supported by the National Natural Science Foundation of China (Grant Nos. 12074165, 12104204, and 11674150), Guangdong Provincial Key Laboratory of Advanced Thermoelectric Materials and Device Physics (Grant No. 2024B1212010001),and the Shenzhen Science and Technology Program (Grant No. 20231117151322001).